\documentclass[10pt,a4paper,twocolumn]{article}

\pdfoutput=1 
\input{Style.dat}
\usepackage{graphicx}
\usepackage{mathtools}
\usepackage{siunitx}
\usepackage[font=small]{subcaption}
\usepackage{nameref}
\usepackage{amsmath} 
\usepackage{chemformula}
\newcommand{\V}[1]{\textup{#1}}
\newcommand\SpecIn{\V{i}}  
\newcommand\LatIn{\V{j}} 
\newcommand\Inlet{\V{in}} 
\newcommand\MolIn{\V{mol}} 

\begin{document}

\title{\large \bfseries Enhanced Modeling of Back-Mixing in Chemical Reactor Networks}
\author[1]{L. Gossel\thanks{Corresponding author: gossel@mma.tu-darmstadt.de}}
\author[1]{M. Fricke}
\author[1]{D. Bothe}
\affil[1]{Institute for Mathematical Modeling and Analysis, Technical University of Darmstadt, Darmstadt, Germany}

\date{} 

\begin{abstract}
\noindent
Chemical reactor networks (CRNs) enable simulations of combustion reactors with detailed chemical kinetics and strongly simplified flow structure. In this paper, the implementation of a reactor component with 
less idealized flow structure, namely axial dispersion, to the CRN software NetSMOKE is presented. It is shown exemplarily that different flow models can lead to strongly different results in species prediction. However, computational efficiency  
remains an issue for  
this reactor class, which is still to be solved. 

\end{abstract}

\maketitle
\thispagestyle{titlestyle}

\section*{Introduction}\label{sec:intro}
Metal energy carriers gain growing interest in research and society as thinkable concept for a convenient, recyclable, low-emission renewable energy storage system \cite{Bergthorson2015,Debiagi2022}. Holistic development of such energy cycles includes the need to overcome large scale- and complexity gaps, i.e.\ from the single particle level to the economic assessment of a full cycle \cite{Neumann2022}. 
Chemical reactor networks (CRNs) are a promising approach to enable reactor simulations with detailed chemical kinetics by employing a strongly simplified flow structure \cite{Khodayari2020,Trespi}. 
As such, in the context of metal energy carriers, they can portentially become a bridge between detailed analysis of metal combustion and reduction on the one hand, and full system evaluation on the other hand. However, in order to achieve this, the method needs further development, which can make the CRN approach even more convenient for reactor simulations in combustion in general.  \\

So far, two idealized reactor models are mostly used in CRN simulations, namely the two limiting cases of complete mixing (Perfectly Stirred Reactor, PSR) and no mixing at all in the axial direction of a pipe flow (Plug Flow Reactor, PFR).
With this, in many application cases, the main species occuring during combustion, e.g. H2O, CO2 and O2 are well predicted, whereas there is a need for more accurate prediction of minor species such as CO in some cases \cite{Trespi}.\\
In Chemical Engineering literature (see e.g. \cite{Levenspiel1998}), other models than the PFR and PSR limit cases are available, which employ a less idealized flow structure. They account for effects such as non-uniform velocity fields, back-diffusion or incomplete mixing, all of them not being represented in PFR and PSR. Therefore, in the current work, it is investigated if modeling non-ideal mixing 
can be, among others, a key to reach better prediction of species in CRNs while enhancing the understanding of internal flow processes at the same time.\\

\section*{Modeling Non-ideal Mixing in CRNs}\label{Model}
One non-ideal flow model that is widely used for modeling purposes of various problems in the literature \cite{Lee1999,Moreau2017} is the axial dispersion model for pipe flows. 
With the modeling being slightly more complex than for PFR and PSR, it can already represent infinitely many flow states between the PFR and PSR limits (see Section \nameref{RTDsec}). This is why the authors have chosen the model for implementation in the CRN framework NetSMOKE (see Section \nameref{Num}). 
The new reactor  
component 
\textit{Axial Dispersion Reactor (ADR)} 
is modeled by a one-dimensional convection-dispersion equation that reads as follows: 
\begin{equation}
\dfrac{\partial \rho_{\SpecIn}}{\partial t} = - \dfrac{\partial (\rho_{\SpecIn} v)}{\partial z} +D \dfrac{\partial ^2 \rho_{\SpecIn}}{\partial z^2}+ M_{\MolIn,\SpecIn}  \omega_{\SpecIn}(\rho_{1},\rho_{2},...).\label{ADRnonsteadystate}
\end{equation}
Here, \textit{$\rho_{\SpecIn}$} is the partial density of the i-th species, \textit{t} is the time and \textit{z} the axial coordinate, \textit{v}(\textit{t},\textit{z}) is the axial velocity, \textit{D} is the dispersion coefficient, $M_{\MolIn,\SpecIn}$ is the molar mass  and $\omega_{\SpecIn}(\rho_{1},\rho_{2},...)$ the production rate of the i-th species due to reactions. \\

The employed boundary conditions are 
\begin{align}
\dot{M}_{\SpecIn,\Inlet}= (\rho_{\SpecIn}\cdot v)_{|z=0} - D\cdot\dfrac{\partial \rho_{\SpecIn}}{\partial z}\ _{|z=0}, \label{BC1}\\
\dfrac{\partial \rho_{\SpecIn}}{\partial z}\ _{|z=L} =0,\label{BC2}
\end{align}
where the inlet condition corresponds to conservation of species mass on the inlet face 
($\dot{M}_{\SpecIn,\Inlet}$ is the area-specific species mass flow into the reactor)  
and there is zero gradient assumed at the outlet (\textit{z}=\textit{L}). 
Currently, only reactive flows that are purely gaseous are considered, so the mathematical model is closed by the ideal gas equation, i.e.:
\begin{equation}
pV=nRT
\end{equation}
with pressure \textit{p}, reactor volume \textit{V}, molar amount of substance \textit{n}, universal gas constant \textit{R} and temperature \textit{T}. It can be coupled to Equations~\eqref{ADRnonsteadystate} to~\eqref{BC2} by $\dot{m}=A v\rho$ and $\rho= \sum_{i} \rho_{i} = \overline{M_{\textup{mol}}}n/V$. Here, A is the cross section of the ADR, $\dot{m}$ is the total mass flow, $\rho$ the total density and $\overline{M_{\textup{mol}}}$ is the average molar mass. \\

Let us note that the dispersion coefficient \textit{D} describes in general not just mass diffusion, but can be orders of magnitude higher than mass diffusion coefficients if intermediate or strong mixing is modeled. It can in principle take any values between zero (plug flow) and infinity (complete mixing). It is assumed here, that \textit{D} does neither depend on the species nor on \textit{z}. 

In order to classify different levels of dispersion, dimensionless measures can be introduced. For axial dispersion, the measure of choice is the Bodenstein Number Bo defined as \cite{Levenspiel1998}
\begin{equation}
\textup{Bo} = \dfrac{L v}{D}.
\end{equation}
For Bo = 1, the dispersion is as strong as convection, whereas for small Bo the flow exhibits strong mixing, and for large Bo the flow shows almost plug flow behaviour. By introducing dimensionless measures for the time and spatial variable and the velocity, the dimensionless form of Equation~\eqref{ADRnonsteadystate} can be derived (adapted from \cite{Levenspiel1998} for compressible flows):
\begin{align}
\theta=\dfrac{t v}{L},\ \zeta =\dfrac{v t +z}{L},\ u = \dfrac{v A \rho_{\textup{in}}}{\dot{m}}\\
\dfrac{\partial \rho_{\SpecIn}}{\partial \theta} = -\left( \dfrac{\partial \rho_{\SpecIn}}{\partial \zeta} + \dfrac{\rho_{\SpecIn}}{u} \dfrac{\partial u}{\partial \zeta}\right) +\dfrac{1}{Bo} \dfrac{\partial ^2 \rho_{\SpecIn}}{\partial \zeta^2}+ M_{\MolIn,\SpecIn}  \omega_{\SpecIn}.
\end{align}

Currently the authors only consider CRNs in the steady state, where this also holds for every modeling component. 
Thus, the steady state ADR has been implemented, which means that the left-hand side of Equation~\eqref{ADRnonsteadystate} is zero.

\section*{Numerics}\label{Num}
For CRN simulations, the authors use the software framework NetSMOKE, which has been created recently 
within the CRECK Modeling Group at Politecnico di Milano (see \cite{Trespi} and references therein) and is further developed in different groups since then. NetSMOKE builds on the open source framework OpenSMOKE++ \cite{OpenSMOKE,Cuoci2011} 
also developed in the CRECK group, which provides solutions for different reactor models such as PFR and PSR with detailed kinetic mechanisms. NetSMOKE is capable of solving chemical reactor networks built from some of these reactor models. 
Currently, different types of PFR and PSR models are available in NetSMOKE \cite{Trespi}. 
The tool provides different numerical (differential) equation system solvers partially inherited from OpenSMOKE, that are used to solve single reactors and the full network (for details see \cite{Trespi}). 

\begin{figure}[h!]
	\centering
	\includegraphics[width=0.99\columnwidth]{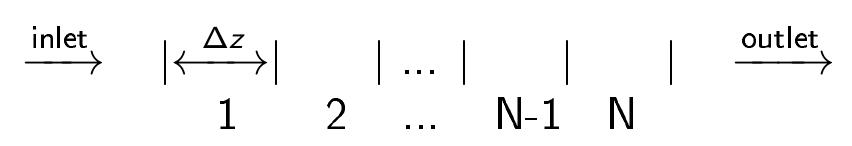}
	\caption{Depiction of the grid on which the ADR equations are discretized for numerical solution. In the current version, the finite volumes are equally spaced and have the length $\Delta z$. The finite volumes are numbered from 1 on the inlet site to N on the outlet site. }
	\label{ADRlat}
\end{figure}

The steady state ADR equations stated in Section~\nameref{Model} constitute 
a boundary value problem for an ordinary differential equation (ODE) of second order. The ODEs are  
discretized  
by the Finite Volume Method \cite{Leveque2002} (see Figure~\ref{ADRlat}):
\vspace{-0.5\baselineskip}
\begin{multline}\label{LeftCell}
\textrm{Inlet cell:}\ 
(0=)\dfrac{\partial \rho_{\SpecIn}^{1}}{\partial t} = \dfrac{\dot{M}_{\SpecIn,\Inlet}-v^{1} \rho_{\SpecIn}^{1}}{\Delta z} + \\D \dfrac{(\rho_{\SpecIn}^{2}-\rho_{\SpecIn}^{1})}{(\Delta z)^{2}}+M_{\MolIn,\SpecIn}\omega_{\SpecIn}
\end{multline}
\vspace{-2\baselineskip}
\begin{multline}\label{MiddleCell}
\textrm{Non-boundary cells:}\\(0=)\dfrac{\partial \rho_{\SpecIn}^{\LatIn}}{\partial t} = \dfrac{v^{\LatIn -1} \rho_{\SpecIn}^{\LatIn -1 }-v^{\LatIn } \rho_{\SpecIn}^{\LatIn }}{\Delta z} +\\ D \dfrac{(\rho_{\SpecIn}^{\LatIn +1}-2\rho_{\SpecIn}^{\LatIn}+\rho_{\SpecIn}^{\LatIn -1})}{(\Delta z)^{2}}+M_{\MolIn,\SpecIn}\omega_{\SpecIn}
\end{multline}
\vspace{-2\baselineskip}
\begin{multline}\label{RightCell}
\textrm{Outlet cell:}\\(0=)\dfrac{\partial \rho_{\SpecIn}^{N}}{\partial t} = \dfrac{v^{N -1} \rho_{\SpecIn}^{N -1 }-v^{N } \rho_{\SpecIn}^{N }}{\Delta z} -\\ D \dfrac{(\rho_{\SpecIn}^{N} -\rho_{\SpecIn}^{N -1})}{(\Delta z)^{2}}+M_{\MolIn,\SpecIn}\omega_{\SpecIn}.
\end{multline}
Here, the upper index denotes the grid cell, whereas the lower index is the species indicator. 
Note that the boundary conditions are already built into the discrete Equations~\eqref{LeftCell} to~\eqref{RightCell} and that the convection term is discretized by an upwind scheme \cite{Versteeg2007}. 

The steady-state algebraic equation system currently can be solved by a nonlinear system solver available in NetSMOKE or pseudo-transiently with an ODE solver. In the latter case,  
time discretization is provided by the ODE solver available in NetSMOKE and OpenSMOKE (see \cite{OpenSMOKE}), and the species-specific partial densities in the inlet stream, $\rho_{\textup{i}}^{\textup{in}}$, are taken as initial value in every cell and for each species, respectively. 

\section*{Model Analysis: Residence Time Distribution}\label{RTDsec}
The ability of the ADR model to represent non-ideal mixing states 
can be seen when considering the residence time distribution (RTD) of the reactor components. The RTD of a certain flow unit (i.e.\ full reactor or a reactor compartment) is the probability distribution for the residence time of a tracer particle passing that unit. The residence time distribution is an important measure in reactor design since it has a significant influence on the progress of chemical reactions in the reactor \cite{Levenspiel1998}. 
Different reactor models have characteristic RTDs and this measure strongly depends on the mixing properties as can be seen when analysing the RTD for ADRs with different strengths of mixing. 

\begin{figure}[h!]
\centering
\vspace{-0.9\baselineskip}
\includegraphics[width=0.99\columnwidth]{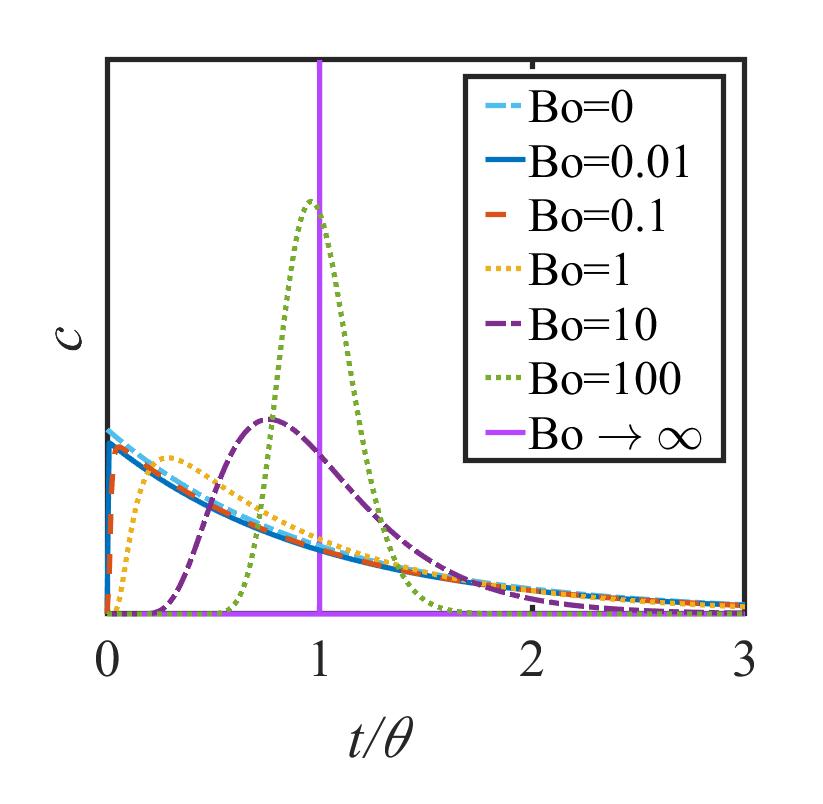}
	\caption{Normed residence time distributions of axial dispersion reactors of length $L$ with varying dispersion (Bo = $ v L/D$, $\theta = L/v$)}
\label{RTDpic}
\end{figure}

In Figure~\ref{RTDpic}, the numerical RTD curves for seven different ADRs including the PFR and PSR limit are shown, assuming a delta-shaped inlet pulse at t = 0. 
The RTD curves have been calculated using a self-written MATLAB \cite{MATLABR2021b} function, as no reactions have to be considered for this. An incompressible velocity field is assumed, an assumption that does not hold in general when reactions are involved, but as the curves shall visualize the effect of mixing behavior the compressibility effects can be neglected here. 

In Figure~\ref{RTDpic}, it can be seen, that the ADR model is able to model RTD behaviors that strongly differ from those of PFR (time delay) and PSR (exponential decay). 
Even for Bodenstein numbers as large as 100, the width of the RTD curve is 
by far non-negligible. 

As the reaction progress depends on the residence time, a residence time distribution strongly differing from the PSR or PFR limit should also lead to different chemical conversion levels at the outlet as would be predicted with the ideal models. 

\section*{Applicability Study: Species Prediction and Computation Time}\label{Appli}
As has been described in the previous sections, one key idea behind the implementation of the ADR is the question, whether the species prediction can be enhanced in some cases by using more realistic flow models. 
On the other hand, as the ADR is to be 
discretized on a mesh, 
which is not the case for PFR and PSR, CPU-time can play a bigger role and might be an obstacle for convenient usage of the model in CRNs. 

Thus, there is a need to assess in which cases the advantages of using an ADR can overcome the additional computational costs and when to use solely PFR, PSR or a network of those to model one compartment. In the present study, the difference in species prediction achievable with the ADR and the respective computation time will be quantified. In particular, the species prediction and computation time of a single ADR will be compared to equivalent networks of PSRs and PFRs. In the scope of this paper, two CRNs are defined as being equivalent if they shall model the same compartment of a full reactor, thus if they have the same total volume, temperature and so on.\\

\begin{figure}[h!]
	\centering
	\includegraphics[width=0.49\textwidth]{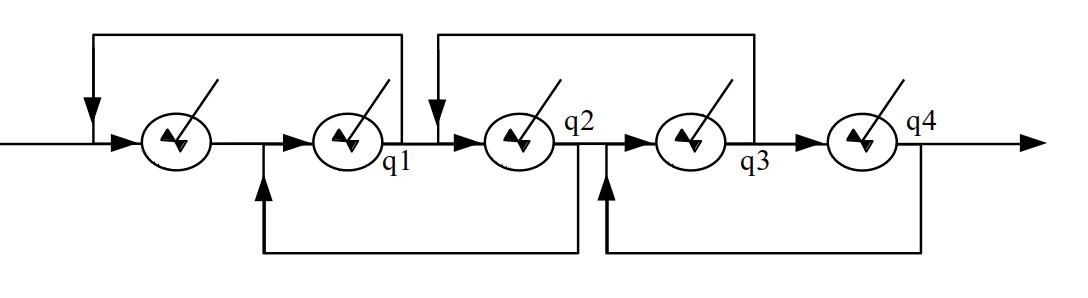}
	\caption{Component representation of the ADR discretized by 5 grid cells. It is solely built from PSRs. Since the splitting ratios q1 to q4 depend on the species concentration and are different for different species, this is however not a general representation of a 5-cell-ADR, which depends on the solution of the ADR. The representation for more grid cells can be obtained by extending the scheme in an analogous manner.
	}
	\label{ADRcanonic}
\end{figure}

Firstly, it will be investigated if there is an ideal, generic representation of the ADR, which is built from PSRs and PFRs. Actually, the ADR discretized by a certain number of grid cells can partially be represented by the network shown in Figure~\ref{ADRcanonic} for the example of five cells, which can be deduced from the finite-volume discretization. However, the splitting ratios q1 to q4 depend on the strength of dispersive back-mixing, thus on the species concentration gradients, so they are in general different for each species. 
Therefore, the ADR solution cannot fully be represented by a combination of PSRs and PFRs. The goal of the present study still is to see how well ADR results can be predicted by simple combinations of PFRs and PSRs which potentially have lower computation times than the ADR. 

The study is structured as follows: 
In a thought experiment, it is assumed that a compartment in a reactor  
is ideally or nearly ideally modeled by an ADR with a certain Bodenstein Number. 
The predicted species and CPU-time is compared to the respective output and CPU-time of an equivalent CRN built solely from PFRs and PSRs, which shall model the same compartment as the single ADR. \\

In an automated procedure, ADRs with Bo $\in \{0.1,1,10\}$ and different numbers of grid cells between 5 and 500 have been tested.\\

\begin{figure}[h!]
\centering
\includegraphics[width=0.99\columnwidth]{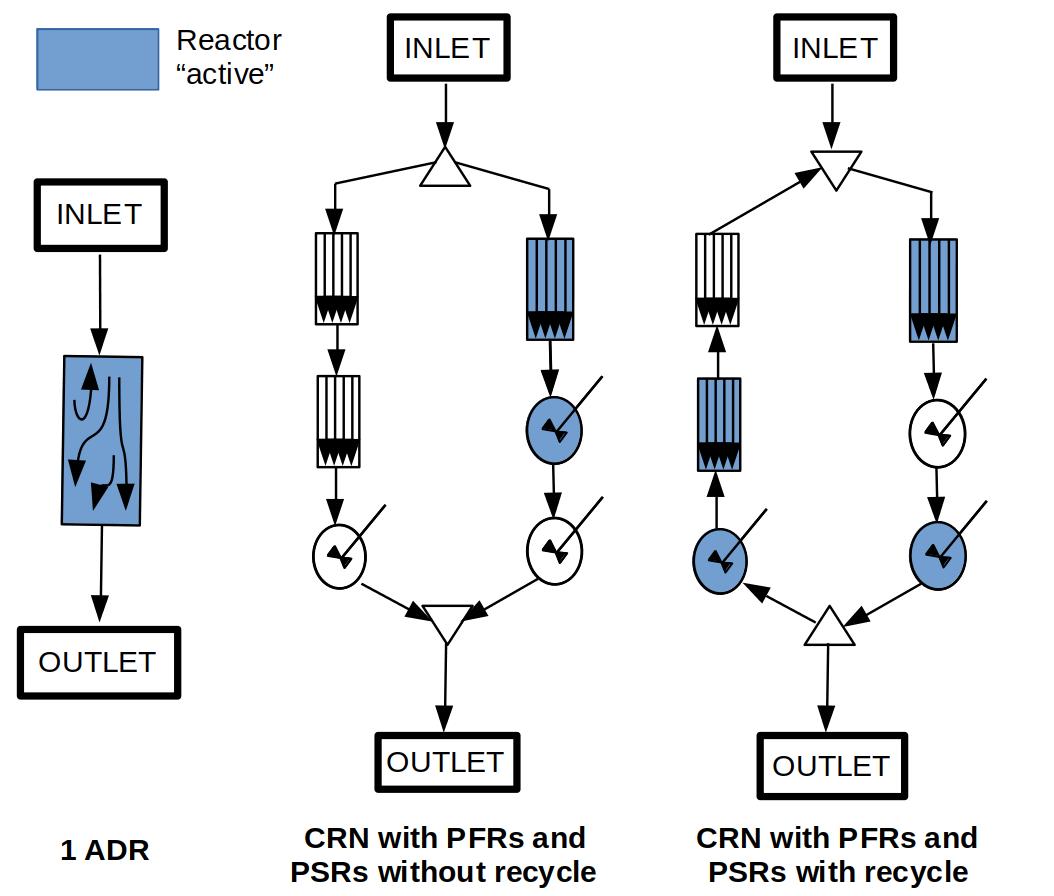}
\caption{CRN templates for the ADR (left) and equivalent CRNs without recycle (middle) and with recycle (right). At the splitting units, half the mass flow enters each outlet stream of the splitter. 
For a more detailed explanation refer to the text. The two example combinations of active reactors shown here refer to CRNs No. 7 and No. 16 in Figure~\ref{FullResultFig}. 
}

\label{ADRTFEX}
\end{figure}

The equivalent CRNs consist of 1-6 reactors in different arrangements, including parallel, series and recycle configurations. 
In Figure~\ref{ADRTFEX}, two CRN templates are shown on which 57 non-redundant equivalent networks are built. One template (middle) is a CRN without recycle, consisting of two parallel branches with three reactors each, and in total three PFRs and three PSRs. For the solution of this network, one reactor component can be solved after the other. The other template (right) is built from the same reactor types, but includes a recycle, so here the system has to be solved iteratively or fully coupled; for more details of CRN solution with recycles we refer to \cite{Trespi}. For each simulation of an equivalent CRN based on these templates, a certain number of reactors between 1 und 6 is considered "active", meaning that the CRN is constructed only from the active reactors. 
This way, by using many\footnote{To prepare the automated procedure more efficiently, a procedure where not all non-redundant combinations are found has been used, so not all non-redundant combinations have been simulated, but many and probably all the relevant ones. } non-redundant combinations of "active" reactors, 57 different equivalent CRNs have been simulated and compared to the ADR (Figure~\ref{ADRTFEX}, left). The total volume of all active reactors from the equivalent network is the same as the ADR volume, so the two CRNs can model the same compartment of a full reactor.

The simulations have been conducted for three different kinetic mechanisms.
The first one is an extremely simplified one used for testing and to gain mathematical insights: It is the reaction \ch{A<=>B + C} 
with Arrhenius type reaction rate. 
The other two are realistic mechanisms, one is the CRECK-
G-2003 \cite{Bagheri2020} gas phase mechanism which is for example used for oxyflame-modeling in \cite{Trespi} and has 114 species and 2000 reactions. The other one is the POLIMI H2CO NOX 1412 mechanism \cite{Ranzi2012}\footnote{Mechanisms can be found on CRECK website: \url{http://creckmodeling.chem.polimi.it/menu-kinetics}} with 32 species and 173 reactions. 
For each mechanism, five different temperatures between \SI{800}{\kelvin} and \SI{2400}{\kelvin} have been employed. The inlet mass compositions for the three cases applied to every ADR and equivalent CRN with the respective kinetic mechanism are presented in Table~\ref{InletTab}. 

\begin{table}[h!]
\centering
\begin{tabular}{l c c c}
Mechanism & Inlet mass & Species & Mass\\
&  flow (kg/s)& & fraction\\ \hline
\ch{A<=>B + C} &  0.0019&A&0.8\\
& &B & 0.2\\ \hline
CRECK-G-2003&  0.0019&O2&0.213\\
& &CH4&0.105\\
& &CO2&0.682\\ \hline
POLIMI H2CO &  0.0001&O2&0.2\\
NOX 1412& &H2&0.2\\
& &N2&0.2\\
& &He&0.2\\
& &Ar&0.2\\ \hline
\end{tabular}
\caption{Inlet mass flow and composition for all ADRs and equivalent CRNs modeled with the respective kinetic mechanisms. }
\label{InletTab}
\end{table}

Note, that the output of any ADR computed at a certain temperature and for a certain kinetic mechanism can be compared to the output of any of the equivalent networks simulated at the same temperature and with the same chemistry. Hence, it can be tested if one of the equivalent networks might represent an ADR with Bo=1, or, if not, maybe the one with Bo=10, without changing the equivalent CRN. 
Consequently, for each temperature and mechanism, theoretically 18 ADRs have been compared to 57 equivalent networks. As computations have not always been successful for all the numbers of grid cells, it has been 90 ADRs for the simple mechanism (and all temperatures), 55 for the POLIMI H2CO NOX 1412 and 40 for the CRECK-
G-2003, in the original study. For the POLIMI H2CO NOX 1412 mechanism at Bo=1 and \SI{1200}{\kelvin}, 46 additional simulations, among them 24 converged ones, have been conducted afterwards to gain a better understanding of grid convergence, so the  
total number of ADRs successfully simulated is 209 leading to 11913 comparable pairs in total. 
Furthermore, ADRs with different numbers of grid cells but equal parameters other than that can be compared within the study. \\

\begin{figure*}[h!]
\centering
\begin{subfigure}[b]{0.49\textwidth}
\centering
\includegraphics[width=\textwidth]{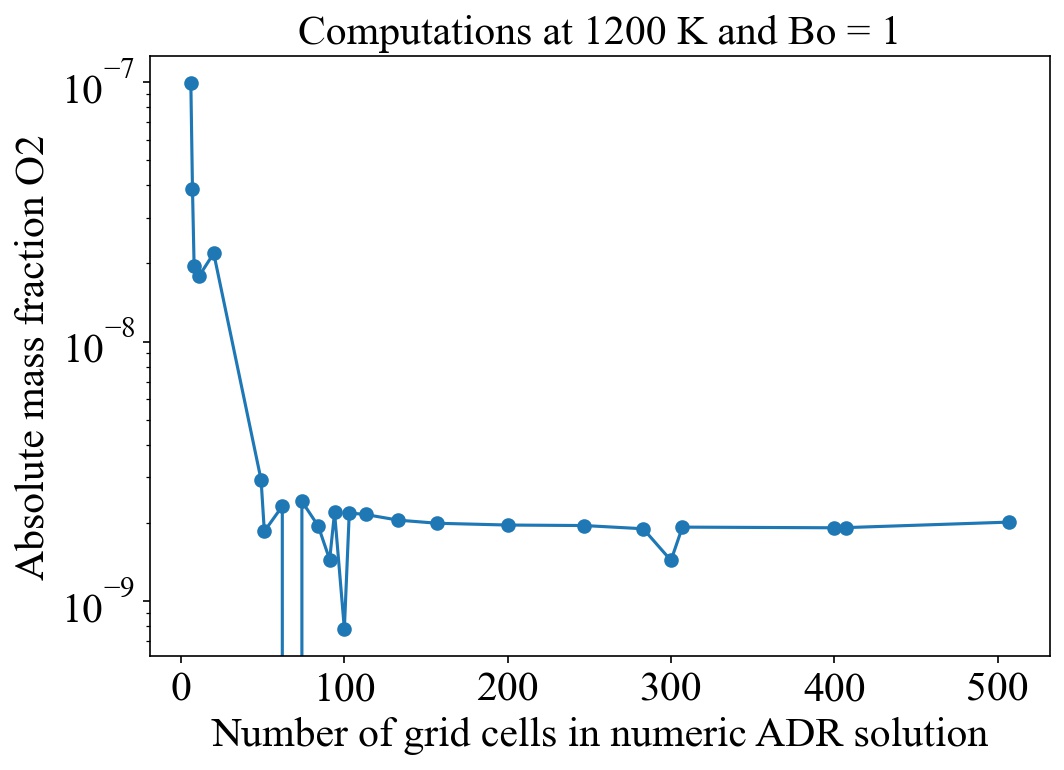}
\caption{Comparison of O$_{2}$ prediction for ADR with different numbers of grid cells.}
\label{ADRLatOutput}
\end{subfigure}
\hfill
\begin{subfigure}[b]{0.49\textwidth}
	\centering
	\includegraphics[width=\textwidth]{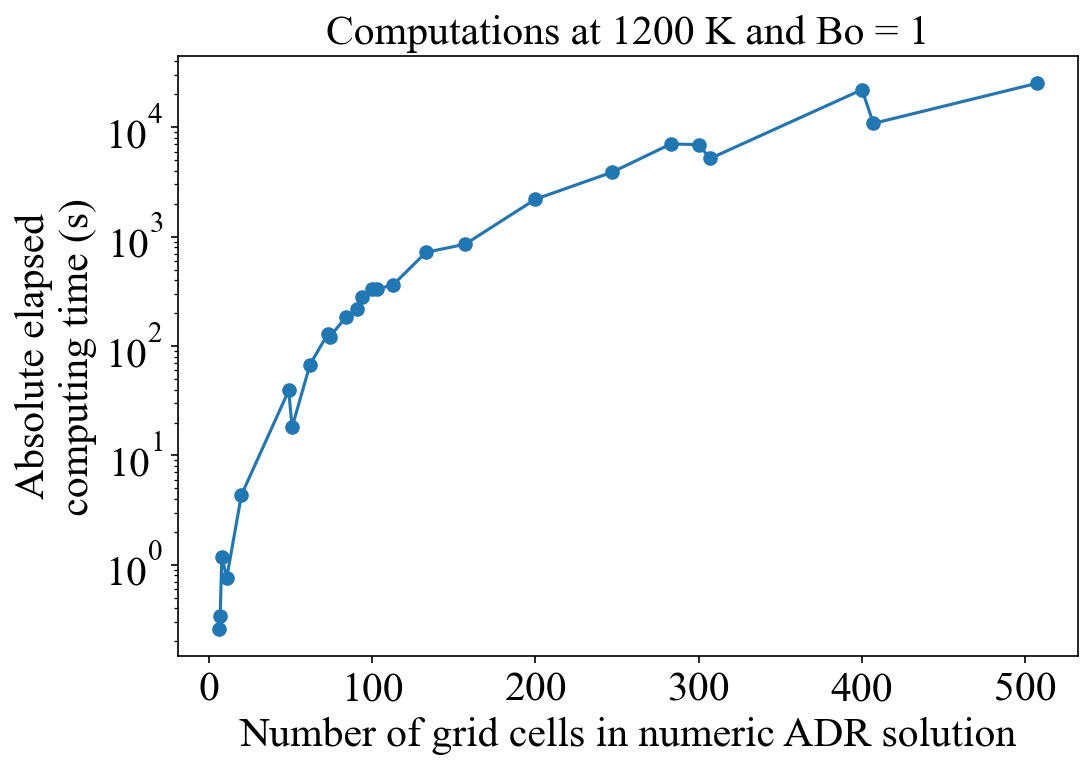}
	\caption{Comparison of computing time for ADR with different numbers of grid cells.}
	\label{ADRLatCPUT}
\end{subfigure}
\begin{subfigure}[b]{0.49\textwidth}
\centering
\includegraphics[width=\textwidth]{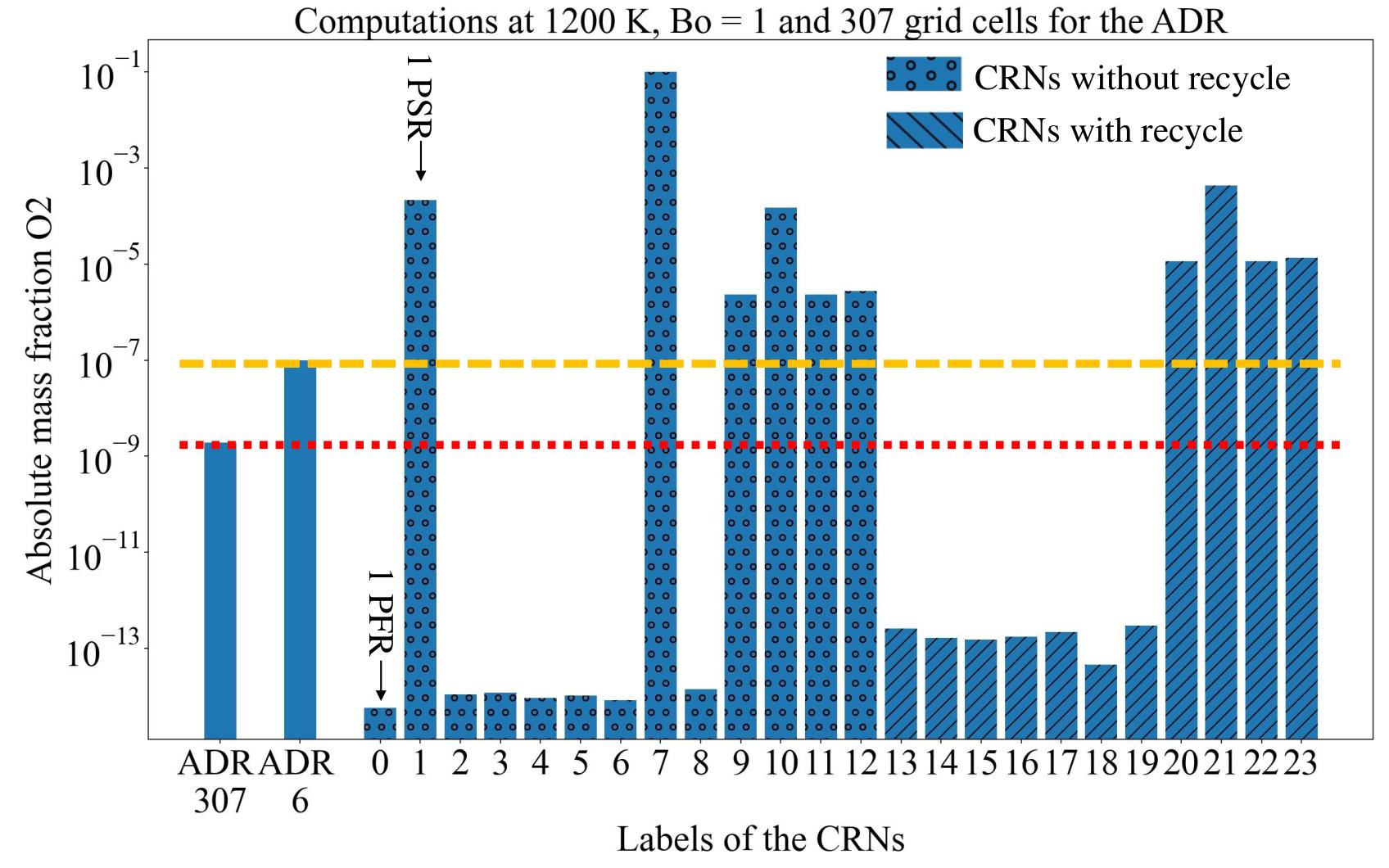}
\caption{O$_{2}$ prediction for two ADRs with 307 and 6 grid cells compared to 23 equivalent networks built from PFRs and PSRs.   
}
\label{RepRNOutput}
\end{subfigure}
\hfill
\begin{subfigure}[b]{0.49\textwidth}
	\centering
	\includegraphics[width=\textwidth]{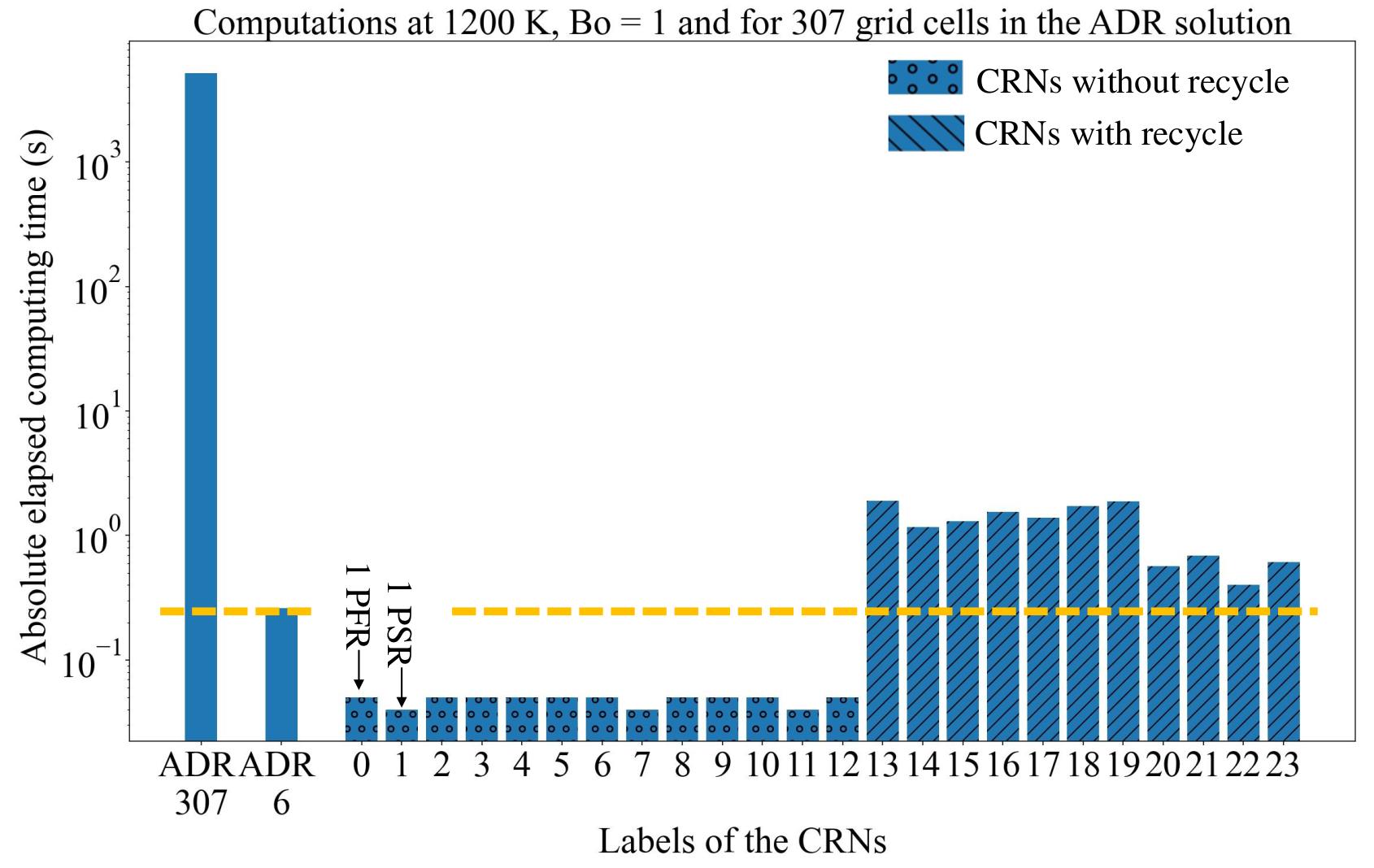}
	\caption{Computation time for two ADRs with 307 and 6 grid cells compared to 23 equivalent networks built from PFRs and PSRs. 
	}
	\label{RepRNCPUT}
\end{subfigure}
\caption{Referring to the study described in this section, results for one ADR with Bo=1 and a realistic kinetic mechanism at \SI{1200}{\kelvin}. In Subfigures~\ref{ADRLatOutput} and~\ref{ADRLatCPUT}, the O$_{2}$ prediction and computation time for ADRs with different numbers of grid cells are shown. In Subfigures~\ref{RepRNOutput} and~\ref{RepRNCPUT}, the results for 307 and 6 grid cells are compared to a selection of equivalent CRNs built from PSRs and PFRs as described in the text and in Figure~\ref{ADRTFEX}. 
In these subfigures, a selection of 23 out of 57 simulations is shown to ease the overview and the numbers are just for enumeration of the cases without any physical meaning. The results for the cases not shown do not lead to other conclusions than the ones shown. For more details refer to the text. }
\label{FullResultFig}
\end{figure*}

Within the scope of this paper, results for Bo=1 and the POLIMI H2CO NOX 1412 mechanism are presented, since they already enable first conclusions. 
The results for computation time and species prediction of O$_{2}$ at \SI{1200}{\kelvin} are presented in Figure~\ref{FullResultFig}. O$_{2}$  has been chosen for presentation since results have shown to be especially insightful for that species. \\

Before studying the results, we focus on the observation 
that the computation of the ADR with realistic kinetics has not always been successful for all the grid cell numbers for which simulations have been conducted. 
This is due to the tremendous number of iterations 
required by 
the nonlinear system solver, which has not been able to converge in certain cases. Although this happens the more often, the higher the number of grid cells becomes, it is still notable that this has also occured for very low grid cell numbers. I.e., it has been observed that the ADR with 5 cells has not converged, but the ones with 6,  7, and 8 cells have. And the ADR with 247 cells has converged, in contrast to the ones tested for 257 and 267 cells. Due to this observations, for the ADR with Bo=1 and the POLIMI H2CO NOX 1412 mechanism, many grid cell numbers have been tested additionally to the ones from the original study. The maximum grid cell number tested is 507 and solution has been successful there. 
The results of all the 26 converged simulations, which have been conducted for the POLIMI H2CO NOX 1412 mechanism and Bo = 1 at \SI{1200}{\kelvin} are shown in  Figures~\ref{ADRLatOutput} and ~\ref{ADRLatCPUT}. 

There, the O$_{2}$ prediction and respective computation time for different numbers of grid cells of the ADR are shown. 
For the O$_{2}$ prediction, it can be seen that grid convergence is reached at around 150 grid cells, except an outlier peak at 300 cells. For simulations between 60 and 100 grid cells, only few simulations have exited successfully and the results undergo strong fluctuations. This behavior needs to be further studied in the future. \\

When the computation time is considered, it becomes clear that the difference in computation time between the ADRs with 6 and hundreds of grid cells is tremendous, being less than a second 
for 6, 14 minutes for 157 and almost two hours for 300 cells. 
Considering that for simulations of real application cases, one ADR often would be only one component of a reactor network to be solved (and there could be more than one ADR for more than one component with non-ideal flow behavior), the computation time can indeed become unfeasible if the time advantage of CRNs compared to CFDs is to be preserved. This holds especially for networks including cyclic structures, where it is impossible to solve one reactor component after the other. However, it is worth taking a look on how the ADR species predictions differ from the ones with other flow models. \linebreak
\phantom{blabla}\linebreak
\indent In Figure~\ref{RepRNOutput}, the O$_{2}$ prediction for the ADRs with 6 and 307 grid cells are shown together with the results of 23 out the 57\footnote{The selection of 23 out of 57 has been done to preserve overview in Figures~\ref{RepRNOutput} and~\ref{RepRNCPUT}, conclusions are the same for the results not shown.} simulated equivalent CRNs. 
It is observed that there is a huge difference in O$_{2}$ prediction between the PSR and PFR results. Whereas the mass fraction is in the order of $10^{-4}$ for the PSR, in the PFR, O$_{2}$ is already completely consumed. The results visible are below the numeric accuracy of NetSMOKE and actually a further study of the PFR profile showed that the reaction occurs in the first \SI{e-5}{\meter} of the reactor, compared to a length of \SI{0.2}{\meter} for the compartment considered here. 
The results for the converged ADR are in the order of 10$^{-9}$, which is still within the numeric accuracy for ADR solution in NetSMOKE. Thus, the ADR prediction is orders of magnitude away from both, the PFR and PSR predictions. Apparently, also the results of all the equivalent CRNs tested are orders of magnitude higher or lower than the ADR results. 

The same holds for the ADR discretized by 6 grid cells even though the results for that ADR are two orders of magnitude higher than for the grid-converged ADR. 
In Figure~\ref{RepRNCPUT}, the computation time of the two ADRs and 23 equivalent CRNs are presented. 
Apparently, in contrast to the converged ADR, the 6-cell-ADR is computational feasible with even lower computation time than the equivalent CRNs with recycle. \\

Overall, it becomes visible that employing a non-ideal flow model as the ADR can indeed make a huge difference in species prediction. Of course, these results depend on the species, kinetic mechanism and reactor considered. For some other species than O$_{2}$ the effect is not as strong, which has also been observed in the current study. The applied flow model is not crucial for all species, but the relevant insight is that there can in general be species for which the flow model is decisive for prediction. Please also note that for the applications for which the kinetic mechanisms, which have been employed here, have been used originally \cite{Trespi,Ranzi2012}, O$_{2}$ is probably not a critical species and is for example stabilized by a cold co-flow. However, this does not affect the conclusion that species conversion can strongly be affected by the mixing quality, depending on the relation of residence time to reaction speed. 

Still, there are further steps to be completed, to make less-idealized flow models as the ADR usable for application. Firstly, all feasible opportunities to reduce computation time by adapting the numerical solution procedure will be tested, with the goal of making the grid-converged ADR applicable. Secondly, it may be further considered to apply a "discrete", thus not grid-converged ADR model, as it can still model non-ideal mixing and has appeared to be computationally feasible. However, this is not as straight-forward in modeling as the grid-converged ADR and more investigation is needed.

\section*{Conclusions}

In this paper, the implementation of an axial dispersion reactor model into the state-of-the-art CRN solver framework NetSMOKE is presented and its applicability is evaluated within a broad parameter study. 
This has revealed that applying a less-ideal flow model can lead to strongly different predictions for some species, compared to the fully idealized PFR and PSR models and also combinations of those. Hence, in compartments with non-ideal flow behavior, employing an ADR can lead to much more accurate predictions for some species.  Nevertheless, computation of the grid-converged ADR is not feasible within CRNs with the current implementation and numeric enhancements have to be pursued. Moreover, alternative models for less-idealized flow will probably need to be considered. Apart from this, further steps also include the validation of the ADR within real application cases.

\section*{Acknowledgements}

The authors would like to thank Prof. Tiziano Faravelli and Prof. Alessandro Stagni for fruitful discussion and support and the whole CRECK modeling group for providing the NetSMOKE framework and the intense collaboration. 

Funded by the Hessian Ministry of Higher Education, Research, Science and the Arts - cluster project Clean Circles.

\bibliographystyle{bibstyle}
\bibliography{bibfile}

\end{document}